# A stellar occultation by Vanth, a satellite of (90482) Orcus

A.A. Sickafoose[a,b], A.S. Bosh[b,c], S.E. Levine[b,c], C.A. Zuluaga[b], A. Genade[a,d], K. Schindler[e,f],

T.A. Lister[g], M.J. Person[b]

[a]*South African Astronomical Observatory, 1 Observatory Road, Cape Town, 7925, South Africa; amanda@saao.ac.za, genade@saao.ac.za*

[b]*Department of Earth, Atmospheric, and Planetary Sciences, Massachusetts Institute of Technology, 77 Massachusetts Ave. Cambridge, MA 02139, U.S.A.; asbosh@mit.edu, czuluaga@mit.edu, mjperson@mit.edu*

[c]*Lowell Observatory, 1400 West Mars Hill Road, Flagstaff, AZ 86001, U.S.A; sel@lowell.edu*

[d]*Department of Astronomy, University of Cape Town, Rondebosch, Cape Town, 7701, South Africa*

[e]*Deutsches SOFIA Institut, Universität Stuttgart, Pfaffenwaldring 29, 70569 Stuttgart, Germany; schindler@dsi.uni-stuttgart.de*

[f]*SOFIA Science Center, NASA Ames Research Center, Mail Stop N211-1, Moffett Field, CA 94035, U.S.A.*

[g]*Las Cumbres Observatory, 6740 Cortona Drive Ste. 102, Goleta, CA 93117, U.S.A.; tlister@lco.global*

Manuscript pages: 37
Figures: 12
Tables: 3
Research data: 3 tables

**Corresponding author:** Amanda A. Sickafoose
SAAO
P.O. Box 9
Observatory, 0735, South Africa
amanda@saao.ac.za
+27(0)21.460.6294

**Abstract:** A stellar occultation by the large trans-Neptunian object (90482) Orcus was predicted to occur on 2017 March 07. Observations were made at five sites in North and South America: the 0.6-m Astronomical Telescope of the University of Stuttgart (ATUS) at Sierra Remote Observatories, California; Las Cumbres Observatory's 1-m telescope at McDonald Observatory, Fort Davis, Texas (ELP); NASA's 3-m InfraRed Telescope Facility (IRTF) on Mauna Kea, Hawai'i; the 0.6-m Southeastern Association for Research in Astronomy (SARA-CT) telescope at Cerro Tololo, Chile; and the 4.1-m Southern Astrophysical Research (SOAR) telescope on Cerro Pachón, Chile. High-speed, visible-wavelength images were taken at all sites, in addition to simultaneous *K*-band images at the IRTF. Solid-body occultations were observed from two sites. Post-event reconstruction suggested an occultation of two different stars observed from two different sites. Follow-up, speckle imaging revealed a second star, which verified that the occulting body in both cases was Orcus' satellite, Vanth. The two single-chord detections, with an anomalously large timing delay in one chord, have lengths of 291±125 km and 434.4±2.4 km. The observations, combined with a non-detection at a nearby site, allow a tight constraint of 443±10 km to be placed on Vanth's size (assuming it is spherical). A 3-$\sigma$ upper limit of 1-4 µbar (depending on constituent) is found for a global Vanth atmosphere. The immersion and emersion profiles are slightly different, with atmospheric constraints 40% higher on immersion than on emersion. No rings or other material were detected within ten thousand kms of Vanth, and beyond 8010 km from Orcus, to the tightest optical depth limit of ~0.1 at ~5 km scale. The occultation probed as close as 5040 km from Orcus, placing an optical depth limit of ~0.3 at ~5 km scale on any encircling material at that distance.

1. Introduction

Stellar occultations provide the most accurate ground-based method for studying trans-Neptunian objects (TNOs), small bodies in the outer Solar System. TNOs are thought to be primitive remnants, better characterization of which can increase understanding of Solar System formation and evolution. The largest TNOs are a few thousand kilometers in diameter. Those sizes correspond to a maximum of a few tenths of arcseconds in angular size when viewed from the Earth. Stellar occultations can be employed to measure sizes of TNOs to an accuracy of a few km, to detect and characterize atmospheres down to nanobar levels, and to discover jets, rings, or satellites (e.g. Benedetti-Rossi et al., 2016; Bosh et al., 2015; Braga-Ribas et al., 2014; Elliot et al., 2010; Gulbis et al., 2006; Millis et al., 1993; Ortiz et al., 2017; Person et al., 2008; Schindler et al., 2017; Sicardy et al., 2011; Young et al., 2008).

A stellar occultation by the relatively large TNO Orcus was predicted to be viewable from the Pacific Ocean and the Americas on 2017 March 07. Orcus is a Plutino, in the 3:2 orbital resonance with Neptune, with relatively high eccentricity and inclination ($a$=39.3 AU, $e$=0.22, $i$=20.5º). Orcus has one known satellite, Vanth, which revolves in a nearly face-on, circular orbit. Vanth's orbit has a period of 9.54 days and a semimajor axis of 8983±26 km or 9030±89 km (Brown et al., 2010; Carry et al., 2011), where the latter result incorporates the astrometric data from the former but could not remove the ambiguity in the orbital inclination. The sizes of Orcus and Vanth have been derived from thermal measurements to be approximately 900 km and 280 km (Brown et al., 2010) and 917±25 km and 276±17 km (Fornasier et al., 2013). The albedo of the system is bright, at an estimated 0.28±0.04 (Brown et al., 2010), $0.27^{+0.07}_{-0.05}$ (Lim et al., 2010), or $0.23^{+0.02}_{-0.01}$ (Fornasier et al., 2013). The bulk density has been reported as 1.5±0.3 g/cm$^3$ or $1.53^{+0.15}_{-0.13}$ g/cm$^3$ (Brown et al., 2010; Fornasier et al., 2013).

Orcus' light curve shows very little change, < 0.04 mag (Galiazzo et al., 2016; Sheppard, 2007; Thirouin et al., 2010). The flatness of the light curve could be due to a spherical shape with little albedo variation, exceptionally low rotation rate, or the viewing geometry being pole-on. Spectral observations of Orcus are dominated by water ice (Lim et al., 2010) with upper limits of 2% methane and 5% ethane, and a feature at 2.2 microns that suggests the presence of ammonia (Barucci et al., 2008; Carry et al., 2011; de Bergh et al., 2005; Delsanti et al., 2010; Fornasier et al., 2004; Trujillo et al., 2005). Hubble Space Telescope data show that Vanth is more red than Orcus in both visible and near-infrared wavelengths, a color disparity that is unusual for TNOs and their satellites (Brown et al., 2010). Orcus is not likely to have retained enough volatiles to support an atmosphere (Schaller and Brown, 2007); however, the compositional combination could hint at a surface renewal mechanism such as cryovolcanism (Barucci et al., 2008; Delsanti et al., 2010).

Here, we report observations of a stellar occultation that was predicted for the Orcus-Vanth system as part of our ongoing efforts to track and characterize outer Solar System bodies (e.g. Elliot et al., 2009). The goal of the observations was to place tighter constraints on Orcus' size (and thus better constrain Vanth's size and the albedos, masses, and densities of each object) and to search for any atmosphere or material around Orcus. An occultation was detected from two sites. We find that the predicted star was occulted by Vanth, as viewed from one site. There was another star, unknown in the prediction, which was occulted by Vanth at a second site. The observations are described in Section 2, with details of the data reduction provided in Section 3. Section 4 contains the results, including size, atmospheric, and ring/debris constraints. A discussion is provided in Section 5.

## 2. Observations

The occultation star was 2UCAC 28372162 at α(J2000) = 10:08:18.54, δ(J2000) = −09:40:14.11, as measured before the event from astrometric data taken in 2014 at the 0.6-m Southeastern Association for Research in Astronomy (SARA-CT) telescope at Cerro Tololo, Chile. The stellar magnitudes are $G$=14.3 (Gaia DR1), $B$=14.98, $V$=14.57, $R$=14.44 (NOMAD), $g'$=15.14, $r'$=14.37, $i'$=14.11 (APASS DR9), $J$=12.98, $H$=12.56, and $K$=12.39 (2MASS). The predicted geocentric midtime for the occultation was 2017 March 07 06:56:48±00:05:58 UT. At the time of the event, Orcus was 19.07 apparent visual magnitude and the system was 47.1366 AU from Earth (from the JPL Horizons database). Note that Vanth is approximately 2.5 magnitudes fainter than Orcus (Brown et al., 2010).

Observations were taken at five stations: the 0.6-m Astronomical Telescope of the University of Stuttgart (ATUS) at Sierra Remote Observatories (SRO), California; Las Cumbres Observatory's 1-m telescope (ELP) at McDonald Observatory, Fort Davis, Texas; NASA's 3-m InfraRed Telescope Facility (IRTF) on Mauna Kea, Hawaii; SARA-CT; and the 4.1-m Southern Astrophysical Research telescope (SOAR) on Cerro Pachón, Chile. Parameters for the observations are listed in Table 1 and additional details are provided in the following paragraphs. Figure 1 shows the predicted shadow path of the event and the locations of each station. For reference, the relative velocity of the star with respect to Orcus was 26.316 km/sec at ELP and 26.271 km/sec at the IRTF.

SRO is located northeast of Fresno, California. An Andor iXon DU-888 camera was mounted on the f/7.9 telescope and ran at -80º C, with 2x2 binning (1.13 arcsec/pixel). The SRO instrument setup is similar to the Portable Occultation, Eclipse, and Transit Systems (e.g. POETS: Souza et al., 2006) and was inherited from characterization measurements for the SOFIA telescope (Pfuller, 2016). For these observations, conditions were clear with high humidity (85%), the seeing was 1.2 arcsec, and the equipment was controlled remotely. Data were recorded from 06:23:16.5 to 07:27:33.3 UT. Timing was provided by a Spectrum TM-4 GPS, to accuracy better than 1 millisecond.

**Table 1**. Observing parameters.

| Site | Coordinates and altitude | Telescope diameter (m) | Instrument | Filter[a] | Exposure time (sec) | Deadtime (sec) | Light curve SNR[b] | Detection |
|---|---|---|---|---|---|---|---|---|
| SRO | 37°04′14″N 119°24′45″W 1.41 km | 0.6 | Andor iXon 888 | clear | 0.2 | 0.0034 or 0.792[c] | 15 | no |
| ELP[d] | 30°40′48″N 104°00′54″ W 2.07 km | 1.0 | Finger Lakes MicroLine 4720 | clear | 2.5 | ~3.4[e] | 24 | yes |
| IRTF | 19°49′34″N 155°28′19″ W 4.24 km | 3.0 | MORIS SpeX | none $K$ | 0.1983 7.0 | 0.0017 ~4.2[f] | 41 3 | yes yes |
| SARA-CT | 30°10′20″S 70°47′57″W 2.12 km | 0.6 | Finger Lakes Instrument | clear | 2.0 | ~0.7[g] | 5 | no |
| SOAR | 30°14′17″S 70°44′00″W 2.74 km | 4.1 | POETS | none | 0.199 | 0.001 | 76 | no |

[a] "None" denotes that no filter was used.
[b] Signal-to-noise ratio (SNR) is defined here as that per 10 km, as an indicative scale for size measurement of an airless body. The noise is defined to be the standard deviation of the baseline.
[c] SRO data were taken as a consecutive sequence of 200-frame blocks at 0.2-second integration. The deadtime between frames within each block was 3.4 milliseconds and the average deadtime between blocks was 792 milliseconds.
[d] ELP is the Las Cumbres Observatory code, based on the airport code of the nearest major city, El Paso.
[e] ELP deadtime varied between 1.47 seconds and 16.45 seconds with a median of 3.41 seconds.
[f] SpeX deadtime varied between 4.11 seconds and 4.20 seconds with a median of 4.18 seconds.
[g] SARA deadtime varied between 0.58 seconds and 0.87 seconds with a median of 0.66 seconds.

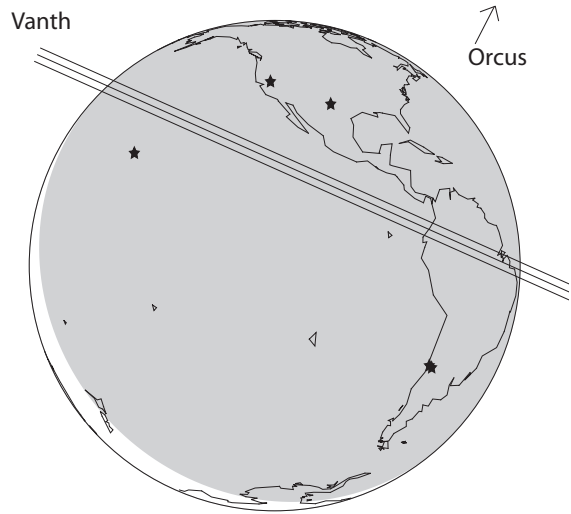

**Figure 1**. Globe of the final prediction for the 07 March 2017 occultation, based on astrometry for the candidate star and the TNO, using the Gaia DR1 network. The locations of the observing stations listed in Table 1 are shown as stars, and the shaded parts of the Earth are locations where the Sun is more than 0º below the horizon. The shadow path for Vanth is shown by solid lines indicating the northern, middle, and southern extent and assuming a size of 280 km. Orcus' shadow is well off the globe to the north.

The Las Cumbres 1-m telescopes are equipped with a standard, wide-field imager that has a 42-sec readout. Through special arrangements, we routinely use the guider cameras on this network of 1-m telescopes for occultation observations. The guider camera is a back-illuminated Finger Lakes Instruments (FLI) MicroLine ML4720. For the ELP (McDonald Obs.) observations reported here, the camera was run at -20ºC, in 2x2 binning (0.674 arcsec/pixel), with 2.5-second exposures and a median readout time of 3.4 seconds with a few longer delays between some frames. The instrument computers are synchronized via NTP to the site GPS timeserver and times are reliable to better than 5 milliseconds. Conditions were clear with 22.7% humidity and seeing of about 1.7 arcseconds. Data were recorded from 06:35:34 UT until 07:19:55 UT.

The MIT Optical Rapid Imaging System (e.g. MORIS: Gulbis et al., 2011) was built specifically for observing stellar occultations from the IRTF. MORIS operates at visible wavelengths and can be used with SpeX (Rayner et al., 2003) to obtain simultaneous near-infrared images or spectra. The MORIS camera was recently upgraded, and it is an Andor iXon DU-897 which was run at −60º C. The data were taken in 1 MHz 16-bit conventional amplifier mode, binned 4x4 (0.46 arcsec/pixel), and using the 0.9-micron SpeX dichroic. Conditions were clear with 30% humidity and seeing was approximately 0.8 arcsec. A few astrometric frames were obtained ~1 to 1.5 hours before and after the event, with the earlier images at airmass > 2. The predicted midtime for Mauna Kea was approximately 07:00 UT, and high-cadence data were taken with MORIS from 06:46:00.0 to 07:22:40.0 UT. Each of the 11000 frames was triggered using a Spectrum TM-4 GPS, to microsecond accuracy. Movie mode for SpeX was not available, so we optimized signal and duty cycle by taking *K*-band images with Guidedog, the slit viewer and guide camera of the spectrograph. The camera was unbinned (0.116 arcsec/pixel) and there were 8 non-destructive reads. Data were taken with SpeX from 06:49:09.9 to 07:16:20.4 UT and the start time of each frame was GPS time-tagged to millisecond accuracy. However, the SpeX timing accuracy per frame is ±0.13 sec as a result of an unknown communication delay between the start selection and the array readout (T. Denault, personal communication).

Data were obtained with SARA-CT using an FLI ProLine PL1001 camera, with a CCD consisting of 1024x1024 24-μm pixels (Keel et al., 2017). A subframe of 512x512 was employed while binning 2x2 (1.21 arcsec/pixel) to minimize readout time. The field of view corresponded to roughly 5x5 arcminutes. Exposures were 2 seconds, with a median readout of 0.66 seconds. The CCD temperature was set to −40ºC and conditions were clear with seeing of 1.6 arcseconds, low wind, and humidity levels around 40% throughout the night. Separated photometry images were taken approximately 5 hours before the occultation at the beginning of the night. High-cadence images were taken from 05:20 UT until 07:40 UT.

Data were taken at SOAR using a POETS instrument built at MIT (Souza et al., 2006). POETS is designed around an e2v CCD-97 optical, frame transfer CCD. The camera is an Andor iXon, model DU-887ECS-BV-9CJ. At setup, the POETS GPS unit (Spectrum Instruments TM-4) completed a full site survey (approximately 4 hours of GPS signal integration) and the derived site latitude, longitude and altitude are given in Table 1. During the observations, the CCD was thermoelectrically cooled to -60°C. The weather conditions during the night were good, mostly clear with light wind and moderate temperatures. Measured seeing was between 0.8 and 0.9 arcseconds. We were able to get images of the occultation star and Orcus when they were well separated, beginning over 4 hours before the occultation, up to about 1 hour before the event. After the event, the field set before we were able to get completely separated images of the star and TNO. For the observations of the occultation, the camera was read out at 1MHz in 16-bit mode using the conventional amplifier. To reduce the readout noise, and increase readout speed, the data were binned 4x4, for an effective pixel scale of 0.194 arcsec/pixel. High speed data were taken with 0.2-second cycle time for 60 minutes between 06:23:00.0 and 07:23:00.0 UT, with each of the 18000 frames triggered by the POETS GPS. Frame times are accurate at the sub-millisecond level. The POETS field of view on SOAR was 25x25 arcseconds. As a result, we were not able to include a comparison star in the field of view.

### 3. Data Reduction

*3.1 Light curve derivations*

For each of the datasets described in Section 2, photometry was carried out to derive normalized light curves. Specific details on derivations for each light curve are provided below, in the order of stations running north to south. The resulting light curves for all stations are plotted in Fig. 2.

The light curve fluxes versus times surrounding the occulted points for ELP, IRTF MORIS, and IRTF SpeX are provided as research data along with this paper. In each case, Orcus and Vanth are so faint as to not be seen in the resulting light curves at mid-occultation. Therefore, separated photometry was not needed in order to determine the zero point of the normalized light curves.

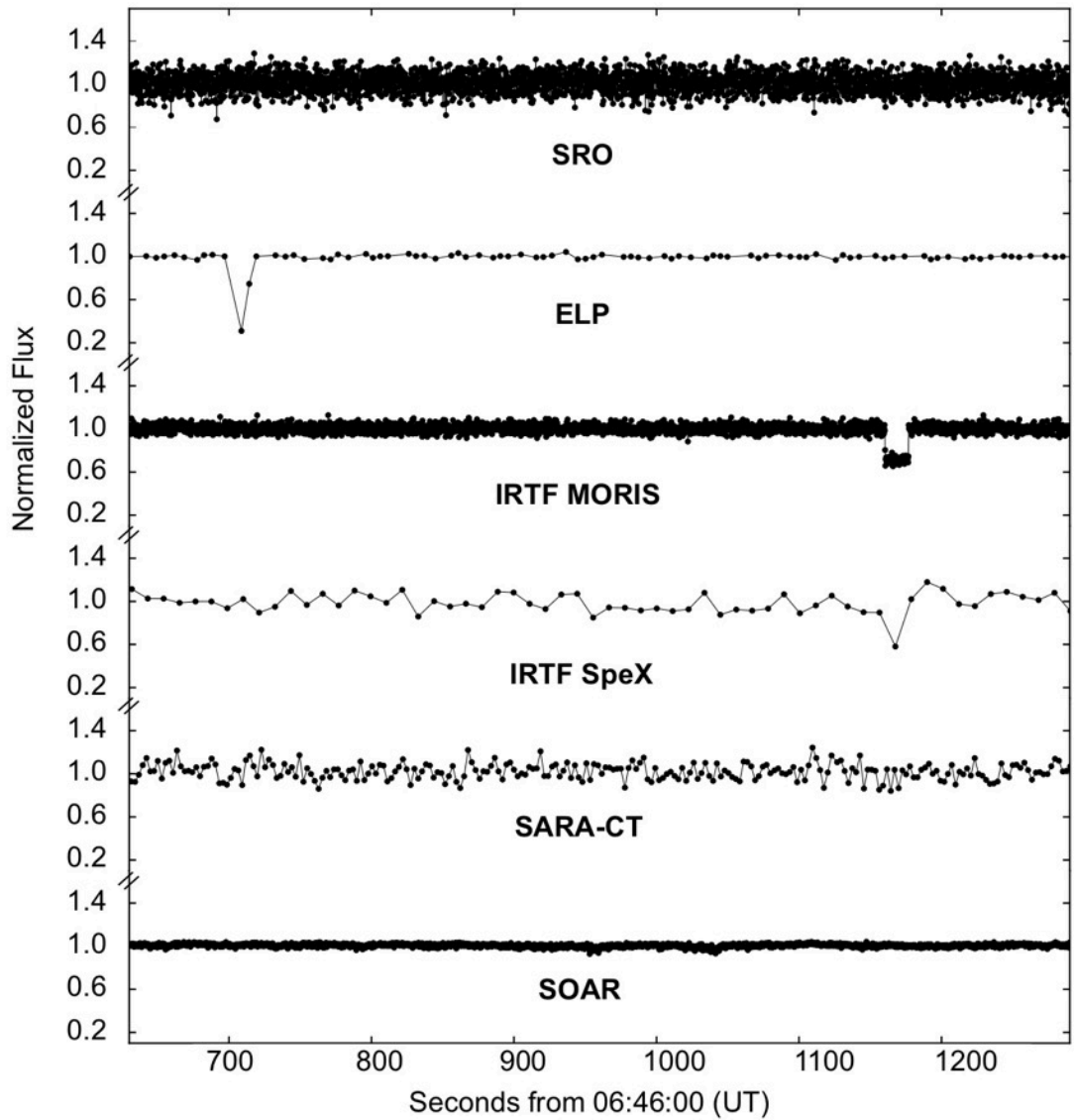

**Figure 2**. Light curves for all stations listed in Table 1. Occultations were detected at two sites, ELP and the IRTF.

Bias-subtracted images were used to generate the SRO light curve. Square-aperture photometry was performed on the target star as well as three comparison stars, with a box size of 8 pixels (9.04 arcsec) returning the best SNR. Two clear-sky regions in the field were used to calculate the average background counts on each frame. The final light curve was calibrated by dividing the target signal by the combined signal of the three comparison stars. Symmetric sections with good SNR were then used to normalize the fullscale flux to 1.

For ELP, bias- and dark-subtracted images were used to generate the light curve. Square-aperture photometry was performed on the target star as well as three comparison stars, with a box size of 8 pixels (5.39 arcsec) returning the best SNR. The average counts from two clear-sky regions were used as the background signal on each frame. A two-point dip signaling the occultation is clearly visible in the target data but not in data for the three comparison stars. The full scale was normalized using symmetric baseline sections. Because the time and spatial resolution were relatively coarse, it cannot be assumed that the flux dropped to zero during the occultation. The background fraction (occulted signal/unocculted signal) is 0.310.

Although taken consecutively, the raw MORIS data at the IRTF consisted of three separate data cubes due to data storage restrictions. All images were stacked and run through a photometry pipeline[1], which was developed for the sister instrument SHOC (Sutherland High-speed Optical Cameras: Coppejans et al., 2013). The header information was updated in the pipeline, including timing for each GPS-triggered frame. Each image was bias corrected and flat fielded, using sky flat images taken during morning twilight. Optimal aperture photometry was performed using PyRAF (a product of the Space Telescope Science Institute, which is operated by AURA for NASA) and DAOPHOT (Stetson, 1987). Optimal apertures for the target and the single comparison star were determined on each frame by using a curve of growth method to maximize the SNR: the smaller of the two optimal apertures was used for both sources on each frame, with an average aperture diameter of 5.2 binned pixels (2.4 arcsec). Sky subtraction was done using the average value in an annulus with inner and outer radii of between one and two binned pixels in diameter larger than the aperture. An obvious occultation, spanning many data points, was apparent in the target and not in the comparison star. This pipeline returns instrumental magnitudes, which were converted into a differential flux ratio of the target divided by the comparison. Normalization of the differential light curve followed by dividing by the mean value of the baseline, outside of the occultation. Based on the flux level during the occultation, the background fraction is 0.710. An extract of the MORIS light curve is provided in Fig. 3 to show the occulted portion in greater detail.

---

[1] http://shoc.saao.ac.za/Pipeline/SHOCpipeline.pdf

For the SpeX data, the *K*-band light curve was derived from raw images. Biases and sky flatfields were obtained: bias-subtracted data returned a light curve of similar profile and similar

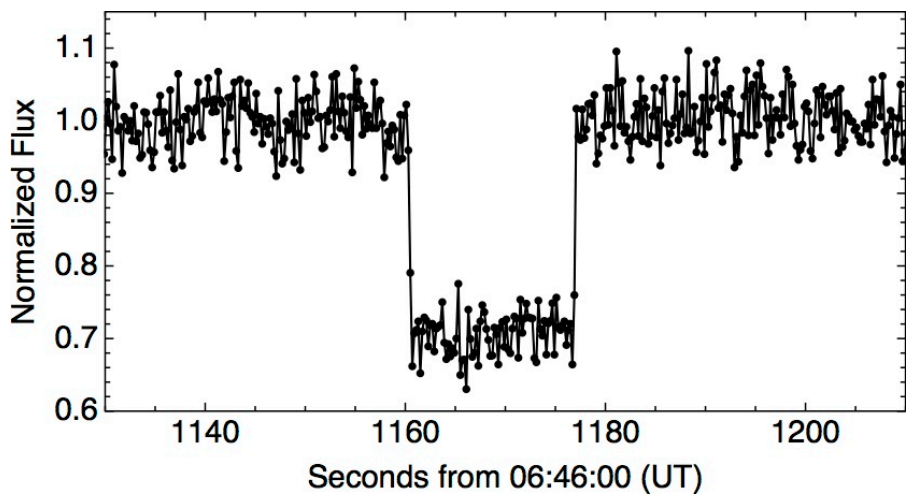

**Figure 3**. Extract of the IRTF MORIS light curve, providing a detailed view of the occulted section.

SNR, flatfielded data returned a light curve of significantly lower SNR. Circular aperture photometry was carried out on both the target star and the comparison, and a 30-pixel diameter aperture (3.5 arcsec) returned the highest SNR. An average from four, hand-selected boxes well outside of the apertures was used as the sky value. Light curves were derived in two ways: (i) by centroiding the target in every frame and (ii) by using a fixed offset from the comparison star as the aperture center for the target. The latter method, using a fixed offset derived from the occultation and comparison star centroids at the start of the occultation dataset, produced the highest SNR. The light curve was calibrated by dividing the target signal by that of the comparison and normalized by dividing by the mean value of the baseline. A single data point was detected during occultation and was contained completely within the occultation as observed by MORIS (see Section 4). For the occulted data point, the background fraction is 0.581. The variation in background fraction between the two instruments is due to an unequal amount of change between visible and NIR flux for the occulted star as compared to the unocculted objects.

Photometry for SARA-CT was derived for the target star as well as the single comparison, with a box size of 8 pixels (9.68 arcsec) returning the best SNR. An average from two separate sky boxes was used for the background value in each frame. The final light curve was calibrated by dividing the target signal by that of the comparison star. Symmetric sections with good SNR were used to normalize the full scale to 1.

For the SOAR data, the light curve was derived directly from the raw data. The only object visible in the occultation images is the combination of the occultation star and the Orcus system. The field was too small to include a usable comparison star. A curve-of-growth analysis was performed on a subset of the images, resulting in selection of a circular aperture of 18 pixels in diameter (0.87 arcseconds). For each frame, the centroid of the combined objects was determined by finding the peak of the thresholded marginals in a box around the center from the previous frame. The aperture was shifted to the new centroid and then the flux in the aperture was summed. The sky background was estimated from two separate apertures, and the mean sky value was subtracted from each object aperture pixel. Figure 2 shows the light curve normalized by the mean value of the whole data set.

*3.2 Measured chords*

Occultation signatures were observed at two sites, ELP and IRTF. The occultation at ELP spanned two data points, the first at lower flux than the second. If the first occulted point did not reach minimum flux (in which a star was completely blocked by an object), immersion and emersion occurred at 708.73±1.25 and 714.21±1.25 sec from 06:46:00 UT. The errors in this case are half of an exposure time, and the corresponding chord length is 5.48±2.50 sec or 144.31±65.80 km. If the first occulted point reached minimum flux, which is the most likely case, immersion and emersion occurred at 702.82±4.66 and 713.88±0.09 sec from 06:46:00 UT. This chord length is 11.06±4.75sec or 291.1±124.9 km. Immersion has a large error due to an unusually large deadtime of 9.32 sec between the last baseline point and the first occulted point (see Table 1 for deadtime statistics). The error on emersion is less than half of the exposure time because we assume instantaneous flux change: we assume the partial-flux emersion data point consists of a fraction of signal at either minimum flux or one. The error on the emersion time thus stems from the noise on the flux, which is the standard deviation in the baseline (0.035).

The Fresnel scale for the event was $\sqrt{\lambda D/2}$ = 1.03 km and 1.97 km, for the visible (600 nm) and *K*-band (2.2 microns), respectively. The stellar diameters of the two stars are 0.73–0.90 km and 0.68–0.79 km projected at the distance of Orcus, based on the equations for giant/supergiant stars from van Belle (1999). The shortest integration time corresponds to 5.25 km, so diffraction broadening and effects from the finite stellar size would be contained within a single data point for all observations.

MORIS at the IRTF detected the most obvious occultation by a solid body. For these data, the error on the flux (standard deviation of the baseline) is 0.034. The occultation chord extends between 1160.47±0.03 sec and 1177.00±0.06 sec from 06:46:00 UT (07:05:20.47 UT to 07:05:37.00 UT). Figure 3 demonstrates that this light curve has a partial-flux data point on immersion (thus the timing error corresponds to noise on the flux for one data point), and the emersion occurred between two exposures (thus the timing error corresponds to noise on the flux for two data points). There is a slightly high data point just before emersion, but it is within the noise of the occulted signal and is assumed to be fully occulted, while the subsequent point is at full flux level. As described above, an instantaneous flux change is assumed. The chord length corresponds to 16.53±0.09 sec duration or 434.39±2.36 km.

*3.3 Event Geometry*

The observed light curves are inconsistent with an occultation of a single star by a single body for two reasons: (i) the chord from SRO was a non-detection; this station lies between the IRTF and ELP chords (Figs. 1 & 2), and (ii) the occulted baseline values of the IRTF and ELP chords are different (Fig. 2) and sum to approximately 1.0. The different depths are most commonly seen in the occultations of multiple stars; as one star is occulted, light from a second star remains in the lower baseline value, raising it above 0. There is uncertainty in the lower baseline value for the ELP light curve. Due to the slow cadence and relatively large read time during which photons were not being collected, we do not know if the lowest detected point (at 06:57:48.73 UT) is the actual bottom of the light curve.

Upon suggestion by a reviewer, we applied for and were granted Fast Turnaround time on the 8.1-m Gemini South telescope to study this occultation star for binarity or duplicity. We used the Differential Speckle Survey Instrument (DSSI, Horch et al., 2009; Howell et al., 2011) in the standard configuration. The DSSI plate scale is 0.011 arcsec/pixel. The data were analyzed by the DSSI team using their standard pipeline, and the resulting image is shown in Fig. 4. The occultation star was found to be double, with a Δm of 0.93 at 692 nm, a separation of 252±2 mas, and a position angle of 316.006 or 136.006 deg (with error ~0.6 deg). Ambiguity of quadrant is common in analysis of speckle data due to autocorrelation analyses. Comparison with our data shows that the 316 deg value fits well, and as a result we use this value for our analyses; the other value does not fit at all.

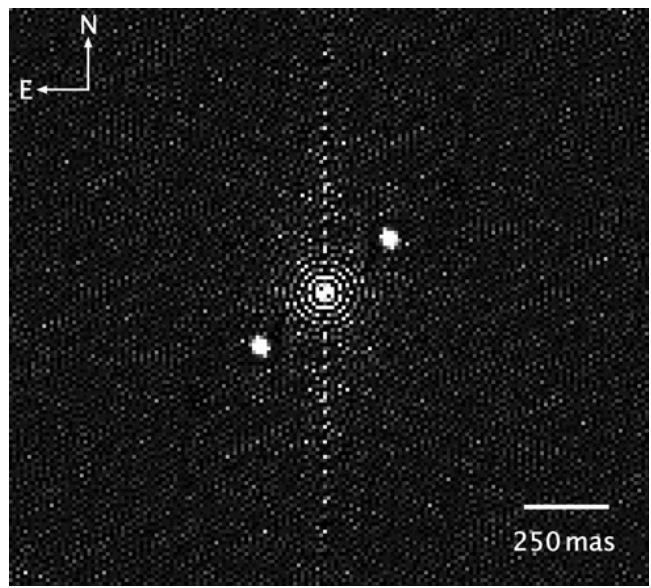

**Figure 4.** Subframe of the reconstructed image from 1000 s of Gemini DSSI data, with the target star in the center. There is a natural quadrant ambiguity in the analysis of speckle data: two secondary stars are shown in the image, while only one is in the correct location. In this case, the correct secondary star is at a position angle of 316 deg (north through east): it is the star on the upper right in the image.

Table 2 contains the observed parameters of this double star. Our original prediction was based on UCAC2; however, the prediction at the time of the event utilized Gaia DR1 (as listed in Table 2). The star position from Gaia DR2 was released after the first submission of this paper. Neither Gaia DR1 or DR2 contained a duplicity flag for this star, and DR2 does not provide proper motion. Note that the observations do not determine whether this is a physical double or simply a projection of two unrelated stars. The newly-determined parameters for stellar position were compared with ephemerides for Orcus and its satellite Vanth. Fitting for occultations of both stellar components by Vanth, and allowing an additional positional offset for the Orcus-Vanth system (see next paragraph), returns excellent agreement with the light curves. The geometry that returns the observed light curves is shown in Fig. 5, with a Vanth occultation by the fainter star at IRTF and by the brighter star at ELP.

To determine the offset of the Orcus system from its ephemeris value, we use the well-sampled IRTF light curve. We assume that the IRTF light curve is central on the body; this is a reasonable assumption as the difference between a diameter and a chord length is small for spherical bodies: even at an impact parameter of 50% of the radius, the chord length is 0.866 of a central chord. With this assumption, we find an offset from the JPL ephemeris (Table 2) of $(\Delta\alpha, \Delta\delta) = (–0.0022, +0.0005)$ arcsec for the Orcus-Vanth system. We then use this system offset, the measured position of the brighter stellar component, and the observatory location for ELP to determine the location of the brighter star's shadow. Based on these parameters and the event geometry, the shadow would pass over ELP with an expected chord length of 165.56 km and an occultation midtime of 06:57:50 UT. These are consistent with observed values. Additionally, these parameters predict that the shadow would not fall over SRO, as was observed. Also in agreement with observations, the shadow from neither star is predicted to pass over SOAR or SARA-CT. The final reconstructed shadow path for the two stars is shown in Fig. 6, with the sky-plane geometry shown in Fig. 7. central chord.

**Table 2.** Occultation parameters.

| | Parameters | Notes |
|---|---|---|
| **Catalog star position** | | |
| Gaia DR1 | $\alpha$(J2000) = 10 08 18.5496 ± 0.004 arcsec<br>$\delta$(J2000) = –09 40 14.245 ± 0.014 arcsec<br>combined mag = 14.24 | |
| Gaia DR2 | $\alpha$(J2000) = 10 08 18.5441 ± 0.0006 arcsec<br>$\delta$(J2000) = –09 40 14.162 ± 0.0008 arcsec<br>combined mag = 14.16 | No proper motion provided |
| **Observed star position** | | |
| double star center of light | $\alpha$(J2000) = 10 08 18.5407 ± 0.0026 arcsec<br>$\delta$(J2000) = –09 40 14.081 ± 0.0037 arcsec<br>combined mag = 14.3 | Source: 16 frames from DCT using Gaia DR2 as reference network |
| double star parameters | separation = 0.252 arcsec<br>separation angle = 316.006 deg<br>$\Delta m$ = 0.93 | Source: DSSI at 692 nm |
| **Deduced star positions** | | |
| star A (brighter) | $\Delta\alpha$ from CoL[a] = +0.0520 arcsec<br>$\Delta\delta$ from CoL = –0.0538 arcsec<br><br>$\alpha$(J2000) = 10 08 18.5443 ± 0.0026 arcsec<br>$\delta$(J2000) = –09 40 14.134 ± 0.0037 arcsec | CoL is observed position |
| star B (fainter) | $\Delta\alpha$ from CoL = –0.1230 arcsec<br>$\Delta\delta$ from CoL = +0.1272 arcsec<br><br>$\alpha$(J2000) = 10 08 18.5324 ± 0.0026 arcsec<br>$\delta$(J2000) = –09 40 13.953 ± 0.0037 arcsec | CoL is observed position |
| **Vanth position** | | |
| offset from Orcus | $\Delta\alpha$ = –0.0445 ± 0.010 arcsec<br>$\Delta\delta$ = –0.2362 ± 0.001 arcsec | Source: M.Brown, personal communication; also Brown (2018) |
| **Orcus position** | | |
| reference ephemeris | JPL28/DE431 | Source: JPL Horizons Ephemeris System |
| offset from ephemeris | $\Delta\alpha$ = –0.0022 arcsec<br>$\Delta\delta$ = +0.0005 arcsec | |

[a] Center of light (CoL).

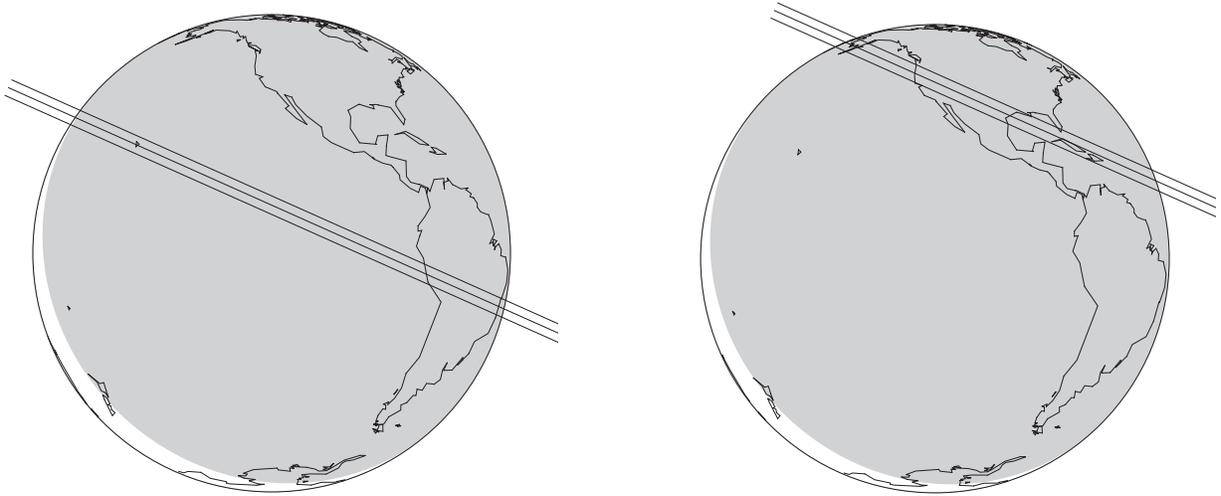

**Figure 5.** Globes for the occultation geometry for Vanth occulting both stars. (*left*) Vanth detection at the IRTF by the fainter star (B). (*right*) Vanth detection at ELP by the brighter star (A). The positions of stars A and B are those listed in Table 2, and Vanth is assumed to have a diameter of 434.4 km, the length of the IRTF chord. The Orcus-Vanth offsets listed in Table 2 are applied to place the IRTF at the centerline at the observed mid-time. The uncertainty in the placement of these occultation tracks is approximately as large as the thickness of one of the track lines.

*3.4 Astrometric data on the night of the event*

Although now superseded by the analysis above, we also collected astrometric images immediately before and after the predicted midtime, with the intention of using these to determine the event geometry. Of the observing stations, SOAR (with a POETS instrument) was the only site that obtained sufficient, well-separated data before and after the predicted midtime. From the IRTF, the event occurred ~1.5 hours after astronomical twilight; therefore, separated data were only obtained afterwards for approximately 1.25 hours. The bulk of the SOAR data were taken one to four hours before the predicted time of the occultation, between 01:40 UT and 05:46 UT. More frames were taken up to 06:12 UT, but these were not used due to blending of the point-spread-functions. Three additional images were obtained after the event, between 08:20 and 08:26 UT. The images were a combination of 30- and 60-second exposures. Thresholded marginals were fit separately for the centroids of the star and Orcus-Vanth, which were both well exposed and had no saturated pixels in all frames. The post-event images were partially blended, and the centroid fitting was done manually to ensure that the process accounted for the blending. For this analysis, Orcus' velocity is assumed to be constant and linear over the duration.

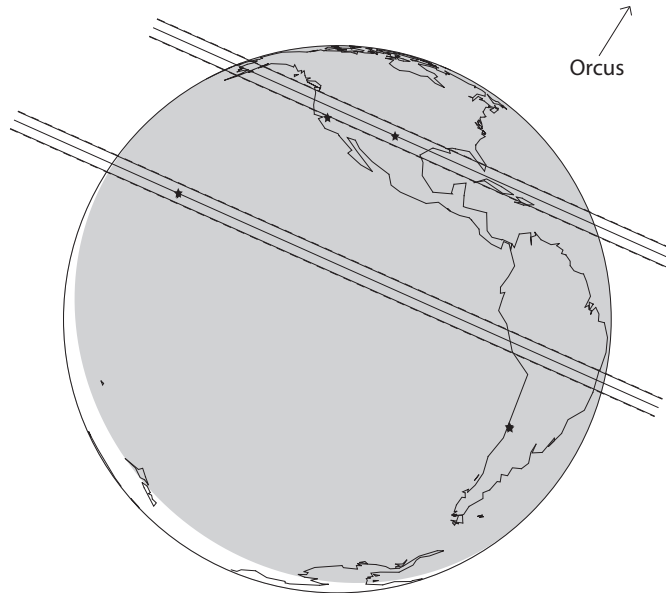

**Figure 6.** Reconstructed shadow paths of the double-star occultation by Vanth, with observing locations indicated by stars. The occultation of the brighter star is the upper path which goes over the southwest US; a chord was detected at ELP in Texas, but not detected at SRO in California. The occultation of the fainter star is the path near the center of the globe; a chord was detected at the IRTF in Hawaii. No chords were detected at SOAR or SARA-CT, both indicated by a single star in Chile. The shadow paths are indicated by three lines (upper limb, center, lower limb), and correspond to a Vanth diameter of 442.5 km (the optimally-determined size in Section 4). The uncertainty for both shadow paths is approximately equal to the width of a single plotted line.

Figure 8 shows the centroids for all the pairs of centers from the astrometric data, shifted to line up the occultation star that is defined to be at position (0,0). A straight line is then fit to the shifted Orcus-system centroids. The distance of the fitted line from the origin is the close approach distance, from which the angular velocity and the time of closest approach are derived. For completeness, this fit was done with and without the three points after the event. The results are very similar. Based only on data taken before the event, the closest approach distance is 0.347 arcseconds at 06:56:03 UT. Including the three points taken after the event, the close approach distance remains 0.347 arcseconds, but the time shifts slightly later to 06:56:45 UT. The predicted mid-time at SOAR was 06:53:51 UT, which is between two and three minutes earlier than the closest approach derived from the astrometric calculations. These results are consistent with the event geometry presented in Section 3.3.

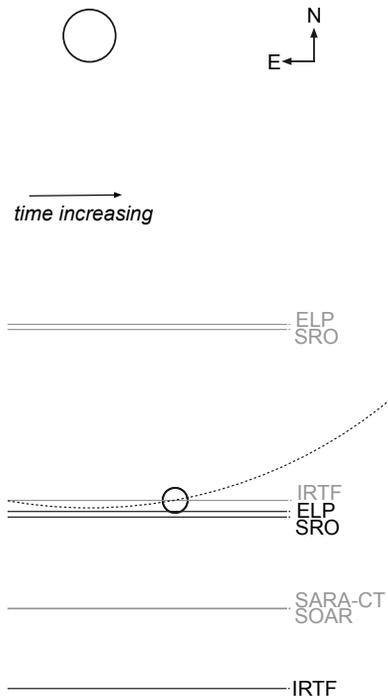

**Figure 7.** Sky-plane geometry for the reconstructed chord paths. Orcus and Vanth (open circles) are assumed to be spherical and are shown to scale, with Vanth's orbit marked by a dotted line. Black lines and labels denote chords for the primary star, gray lines and labels denote chords for the secondary star. Only the chords closest to the bodies are displayed: SARA-CT and SOAR for the primary star are off the figure to the south.

## 4. Results

As presented in Section 3, the primary star was occulted at ELP and a secondary star was occulted at the IRTF. The magnitude of the IRTF star can be estimated from the occultation light curves. For the SpeX data, a single data point was detected during occultation, with midtime 1167.62±0.130 sec from 06:49:00 UT. Even with the errors, that exposure falls completely between the immersion and emersion times observed by the co-mounted MORIS and is thus a fully occulted detection. The background fraction is approximately 0.13 higher in MORIS (visible) data than in SpeX (*K*), which means that the occulted star is significantly brighter in the red than the visible, relative to the unocculted objects (Orcus+Vanth+primary star). The combined stars have magnitudes *V*=14.57 and *K*=12.39. Assuming magnitudes of 19.4 and 22.0 (visible) and 22.0 and 24.1 (*K*) for Orcus and Vanth respectively (Carry et al., 2011), and using the derived background fraction for MORIS (0.710) and SpeX (0.581), the two separate stars have approximate magnitudes 14.95 and 15.90 (visible) and 12.98 and 13.33 (*K*). These magnitudes are approximate because the MORIS response is open, rather than being in a specific visible-wavelength filter. Data from Gemini, in which the two stars are resolved, return a magnitude difference between the two stars of 0.93 at 692 nm and 0.90 at 880 nm. The former of these measurements is close to the assumed visible wavelength of 600 nm (the approximate midpoint of the Andor iXon QE curve) and the Gemini stellar flux difference is similar to that derived from the MORIS light curve.

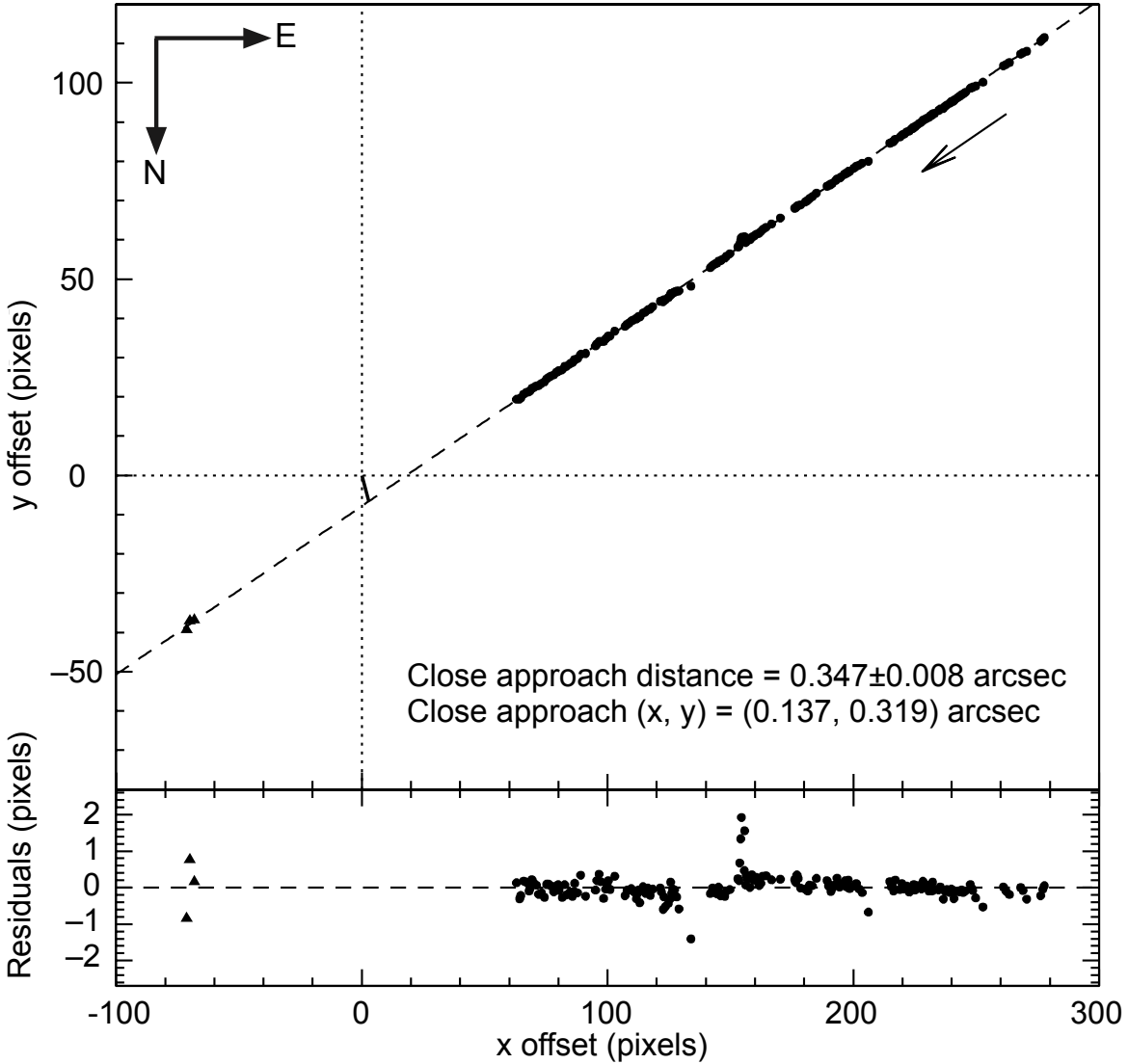

**Figure 8.** Astrometric data from SOAR, with a plate scale of 0.194 arcsec/pixel, used to determine the time and distance of the closest approach between the target star and the Orcus-Vanth system as observed from that site. The origin is the centroid of the star, and the dashed line is a linear fit to the measured centroids of Orcus-Vanth. The arrow indicates increasing time and the solid line extending from the origin indicates closest approach (the line is orthogonal to the fit line, but does not appear so due to the plot aspect ratio). Residuals, the perpendicular values for the data minus the linear fit, are plotted below each point. The dots represent data taken before the predicted midtime, and the triangles are data taken afterwards. A linear fit to all of the data returns the same result as fitting only the data taken before the midtime, within errors.

There are three size constraints placed by the observations. First, MORIS at the IRTF detected the most obvious occultation by a solid body, with a chord length of 434.4±2.4 km. This chord places a lower limit on the diameter of 432 km. Second, the occultation at ELP has a most likely chord length of 291.1±124.9 km. The non-detection at SRO and the length of the chord at ELP place a poorly-constrained upper limit on the size of an occulted spherical object of $329^{+251}_{-162}$ km. Third, with the measured double-star parameters, the projected Vanth offset from Orcus, and the deduced Orcus system ephemeris offset (Table 2), our model for this occultation reproduces the observed light curves and allows a tighter constraint to be placed on the maximum size. Using the event geometry and increasing the size of Vanth until SRO is just outside the shadow path, we find a maximum size of 453 km. Thus, Vanth's diameter is between 432 and 453 km, or 442.5±10.2 km. Because multiple, high-spatial-resolution light curves were not obtained, we cannot fit for a non-spherical shape. Thus, a circular projection of Vanth in the sky plane is assumed.

Although Vanth is small and not expected to have an atmosphere, we can place atmospheric limits by employing a model following French & Gierasch (1976) for a thin atmosphere with diffraction on a solid surface. This model assumes a radially symmetric atmosphere with constant scale height much less than the body's radius, and it is applicable when the scale height is much greater than the Fresnel scale. Fit parameters are listed in Table 3, where $b$ is representative of delta flux at the surface. Immersion and emersion are fit separately, and the MORIS observational wavelength is assumed to be 600 nm. A range of atmospheric scale heights are tested, and the best-fit results are plotted in Fig. 9. The maximum 3-σ bending parameter over this tested range of scale heights is $b=0.34$ on immersion and $b=0.23$ on emersion.

**Table 3.** Vanth atmospheric fit parameters.

| Scale height (km) | Background | Fullscale | Geometric edge (s) | Bending parameter ($b$; $10^{-2}$) | Sum of squares of residuals |
|---|---|---|---|---|---|
| *Immersion* | | | | | |
| 10 | –0.017±0.019 | 0.994±0.017 | 1160.17±0.016 | 0.0003±0.0026 | 1.036 |
| 50 | –0.019±0.017 | 1.026±0.021 | 1160.18±0.018 | 15.0±6.3 | 0.974 |
| 100 | –0.019±0.017 | 1.054±0.031 | 1160.17±0.017 | 14.6±5.9 | 0.970 |
| *Emersion* | | | | | |
| 10 | 0.027±0.017 | 1.001±0.016 | 1176.66±0.023 | 0.002±0.001 | 0.883 |
| 50 | 0.020±0.016 | 1.009±0.021 | 1176.65±0.021 | 4.1±6.5 | 0.885 |
| 100 | 0.020±0.016 | 1.011±0.030 | 1176.65±0.021 | 2.7±6.4 | 0.871 |

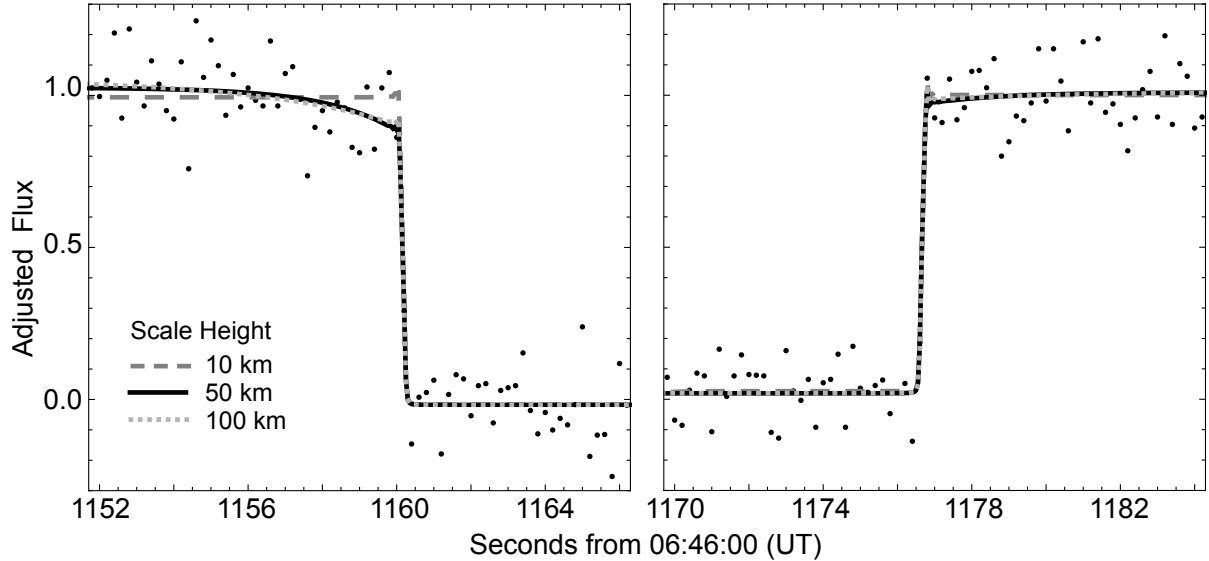

**Figure 9.** Atmospheric model fits to the immersion and emersion data from IRTF MORIS. The best fits are plotted for the parameters shown in Table 3. For the purpose of atmospheric fitting, the flux is rescaled between 1 (full scale) and 0 (during occultation).

Upper limits on an atmosphere can be placed by considering a range of possible constituents, such as those detected in Pluto's atmosphere, on comets, and in Orcus' spectrum (e.g. Bieler et al., 2015; Carry et al., 2011; Meech and Svoren, 2004; Mousis et al., 2013; Womack et al., 2017; Young et al., 2017). Assuming temperature 44 K (upper limit from Barucci et al., 2008), and that Vanth is 443 km in diameter and 1/15 the mass of Orcus, the 3-$\sigma$ upper limits on an atmosphere are between $10$–$50 \times 10^{13}$ cm$^{-3}$ or 1–4 μbar, with constraints on immersion being 40% larger than those on emersion.

Searching for material in the Orcus-Vanth system can be undertaken by considering how close to each of the bodies the light curves probed and testing for symmetry of any features. The two highest-SNR light curves with the closest approach distances are IRTF MORIS and SRO (ELP was closer to both bodies than SRO, but at much lower SNR). The chord geometry is shown in Fig. 7, with the closest approach distances to Orcus for these two light curves plotted in Fig. 10. The standard deviation of the IRTF MORIS baseline is 0.034 flux, with a 3-$\sigma$ upper limit on line-of-sight optical depth of $\tau < 0.11$ at the full spatial resolution of 5.25 km. The SRO standard deviation is 0.087 flux, with a 3-$\sigma$ upper limit of $\tau < 0.30$ at the full spatial resolution of 5.34 km. To look for broader features, a moving average of the data is used, and immersion and emersion are compared to detect any features that are symmetric about the bodies. Figure 11 shows extracts of the light curves close to Vanth's surface, and Fig. 12 shows extracts for the closest approaches to Orcus. Moving averages of 2 and 14 data points are displayed, corresponding to distances of 10.51 km and 73.56 km for IRTF MORIS and 10.69 km and 74.82 km for SRO. These scales were chosen to match those of known rings around minor bodies. For IRTF MORIS, the 3-$\sigma$ upper limits on line-of-sight optical depth at these scales are $\tau < 0.07$ and $\tau < 0.03$, respectively. For SRO, the 3-$\sigma$ upper limits at these scales are $\tau < 0.21$ and $\tau < 0.08$, respectively. There are no statistically significant or symmetric features at any scale, in either light curve. Therefore, no material was detected to these limits at Vanth. No material was detected to these limits around Orcus, beyond closest approach distances of 8010 km (IRTF limits) and 5038 km (SRO limits).

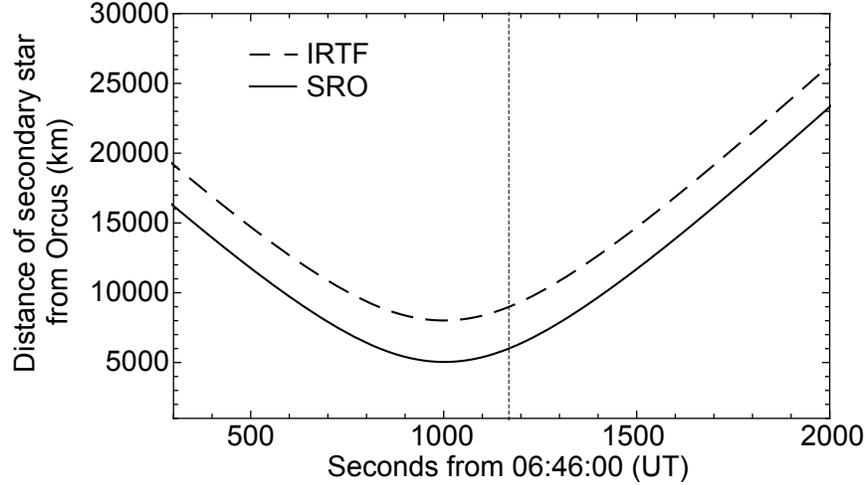

**Figure 10**. Distances of the IRTF and SRO secondary-star chords from the center of Orcus as a function of time. The dotted line represents the midtime of the Vanth occultation. These distances assume that the IRTF is a central chord on Vanth (see Section 3.3).

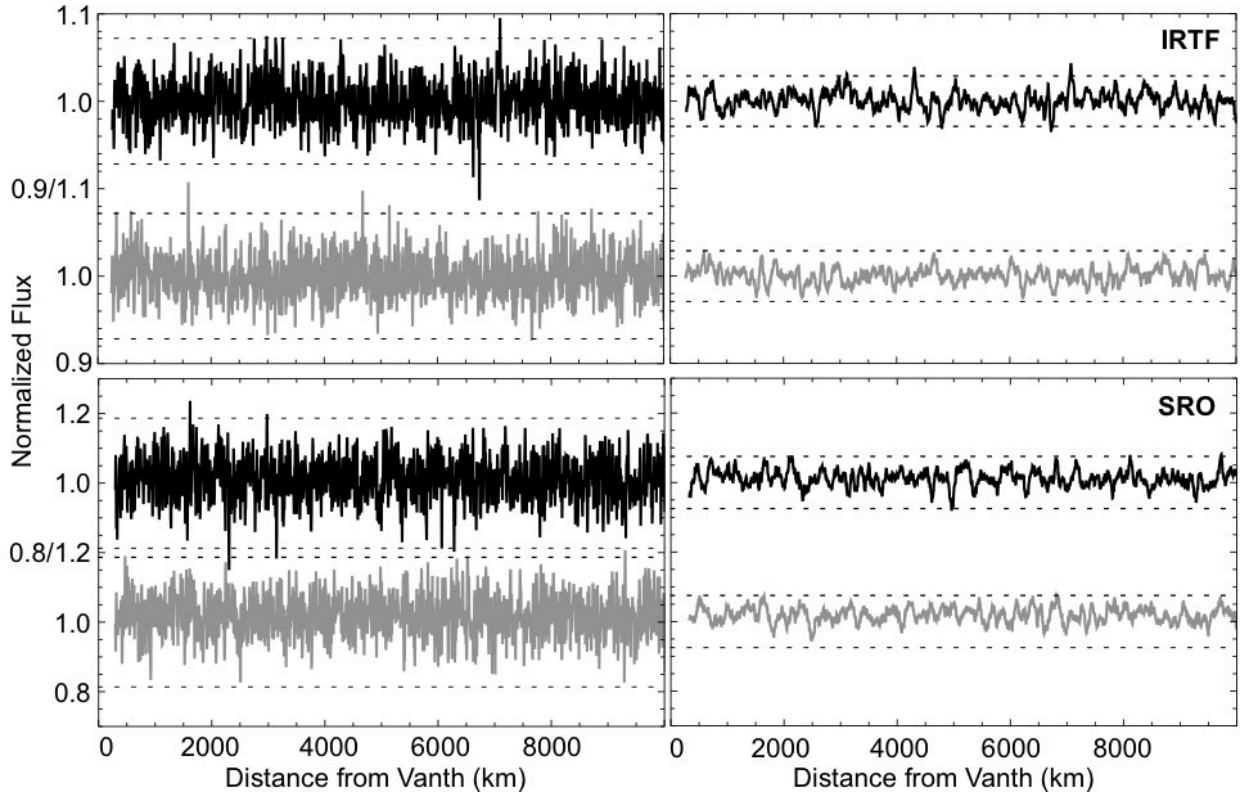

**Figure 11**. Investigation for rings or other surrounding material around Vanth. Data are from (*top*) IRTF MORIS for the secondary star closest approach and (*bottom*) SRO for the primary star closest approach. A moving average is applied to the light curves over scales of approximately (*left*) 11 km and (*right*) 74 km. Immersion data are shown in black and emersion data are plotted in gray. Dashed lines represent three-σ error bars. Azimuthally symmetric features should appear at the same distance from Vanth in both light curves.

## 5. Discussion

We observed a stellar occultation of the 14.95-magnitude primary star by Vanth from Texas and of an approximately 15.90-magnitude secondary star by Vanth from Hawai'i. Assuming it is spherical, we find a diameter for Vanth of between 432–455 km. At over four-hundred km in diameter and an icy object, Vanth is likely to have reached hydrostatic equilibrium (e.g. Duffard et al., 2009). No substantive, global atmosphere was detected on Vanth, to a 3-σ upper limit of approximately 4 μbar for expected atmospheric constituents. In addition, azimuthally symmetric rings/debris are ruled out within ten thousand kms of Vanth and beyond ~5040 km from Orcus, with a 3-σ upper limit on optical depth of $\tau < 0.3$ for narrow features and $\tau < 0.08$ for broad features. Stricter limits are placed on encircling material beyond ~8010 km from Orcus, with a 3-σ upper limit on optical depth of $\tau < 0.11$ for narrow features and $\tau < 0.03$ for broad features.

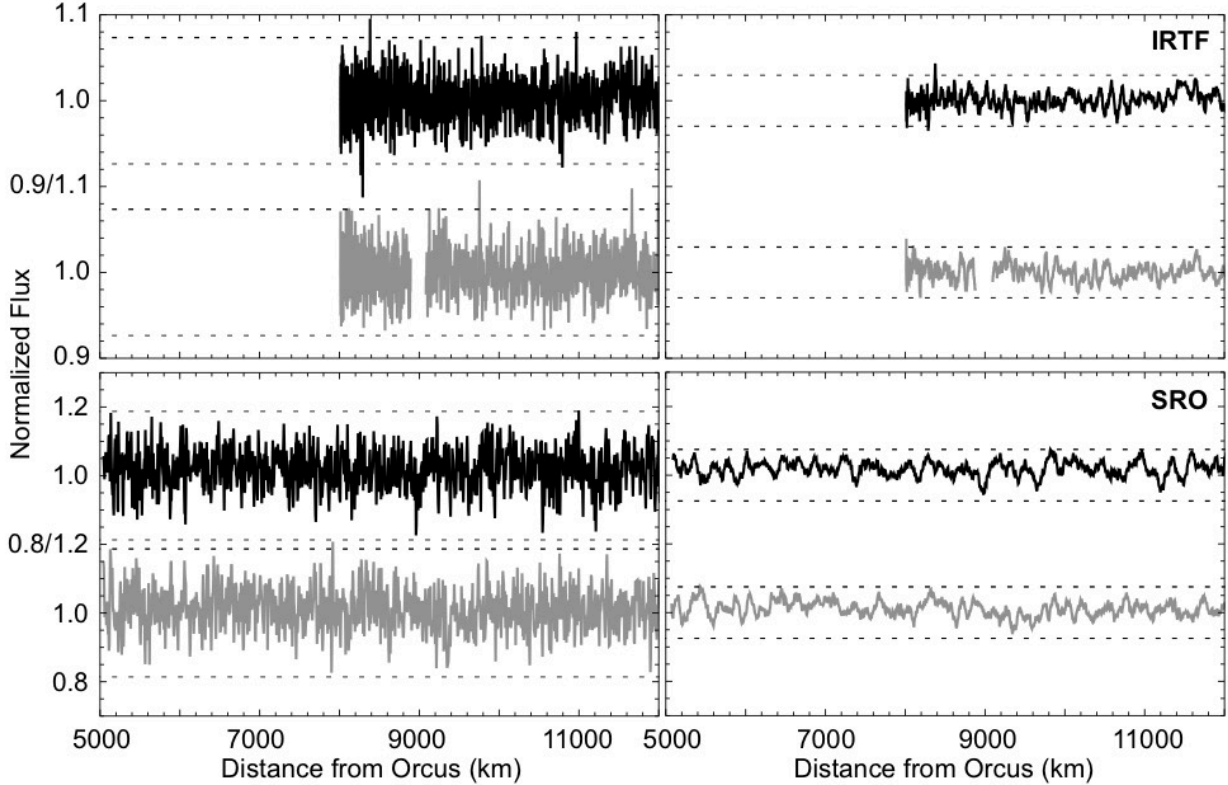

**Figure 12**. Investigation for rings or other surrounding material around Orcus. Data are from (*top*) IRTF MORIS and (*bottom*) SRO, both for the closest approach of the secondary star. A moving average is applied to the light curve over scales of approximately (*left*) 11 km and (*right*) 74 km. Immersion data are shown in black and emersion data are plotted in gray. Dashed lines represent three-σ error bars. Azimuthally symmetric features should appear at the same distance from Orcus in both light curves. The gap in the IRTF Orcus emersion data is due to the Vanth occultation.

The size and albedo of the system have been previously measured, from which estimates for the individual bodies have been derived (Brown et al., 2010; Fornasier et al., 2013; Stansberry et al., 2008). Brown et al. (2010) found that Vanth is significantly more red than Orcus, suggesting the objects could have quite different albedos. Our derived size for a spherical Vanth is significantly larger than previous estimates. However, after this paper was submitted, new measurements were reported from the Atacama Large Millimeter Array, with sizes and albedos of $910^{+50}_{-40}$ km and 475±75 km, and 0.23±0.02 and 0.08±0.02, for Orcus and Vanth, respectively (Brown and Butler, 2018). Our result for Vanth's size is consistent with these new data and places a tighter constraint on the size. Our result does not allow a tighter constraint to be placed on the size of Orcus than that directly measured by Brown and Butler (2018). If the objects have equal density, the mass ratio is between 6 and 12. However, given the difference in size, color, and albedo, Vanth is likely to be less dense than Orcus. Considering the extreme range of likely densities, 0.8 g/cm$^3$ for Vanth and 2 g/cm$^3$ for Orcus, and our optimal size for Vanth, the mass ratio reaches 22.

Models of volatile loss and retention for large TNOs place Orcus firmly in the "All Volatiles Lost" category (Schaller and Brown, 2007). In this model, Vanth, at 443 km diameter and 44 K, falls into the same category: either object would have to be significantly colder (< 30 K) to possibly retain volatiles. Furthermore, Vanth is less than half the size of the smallest satellite known to have a tenuous atmosphere (Dione, Simon et al., 2011). Our upper limit on a tenuous Vanth atmosphere is consistent with this expected lack of volatiles. The lack of an atmospheric signature, and the lack of detection of any encircling material, indicate that no cryovolcanism or outgassing from Vanth was detected in our light curves. Although not likely, the variation between the immersion and emersion profiles allows for possible tenuous, localized, near-surface differences, at levels below those measurable from these data.

The idea that rings or other coma material could exist in the Orcus system is based on recent discoveries of a ring around the dwarf planet Haumea (Ortiz et al., 2017), a pair of rings around the centaur Chariklo (Bérard et al., 2017; Braga-Ribas et al., 2014), and possibly jets, a shell of material, or rings around the centaur Chiron (Bus et al., 1996; Elliot et al., 1995; Ortiz et al., 2015; Ruprecht et al., 2015). The centaur rings are 3–7 km in width and Haumea's ring is

DRAFT OF ACCEPTED MANUSCRIPT                                                                        29

roughly 70-km wide. Orcus falls in between the sizes of these bodies, being smaller than Haumea and larger than the centaurs. The orientation of Orcus is not well known, so we measure line-of-sight optical depth. However, the orbit of Vanth represents a likely ring plane. This nearly face-on orientation implies an almost fully open ring tilt angle and thus the line-of-sight optical depth would be equivalent to the normal optical depth. For comparison, the Haumea ring has optical depth 0.5 (Ortiz et al., 2017), and an upper limit has been placed on material around Pluto of normal optical depth 0.07 for rings a few km in width from ground-based occultations (Throop et al., 2015). Our results rule out any encircling features on similar scales around Vanth and beyond 5040 km from Orcus, with tighter constraints placed beyond 8010 km from Orcus. The detected ring systems around minor bodies are located closer to the surfaces than these chords probed for Orcus: orbital distances from the centers are roughly 2300, 400, and 300 km for Haumea, Chariklo, and Chiron, respectively (Braga-Ribas et al., 2014; Ortiz et al., 2017; Ruprecht et al., 2015). The Roche limit for a fluid satellite at Orcus is roughly 1350 km, assuming densities of 1.5 g/cm$^3$ for Orcus and 0.8 g/cm$^3$ for the satellite. Thus the possibility for rings at Orcus, as well as arcs or non-azimuthally symmetric material, remains. Neither can we rule out cryovolcanism for either Orcus or Vanth, since there could be sparse material below the detection limits, or the observation could have occurred at a time of quiescence.

This is the first two-chord stellar occultation by Orcus' satellite Vanth. A large observational campaign for a predicted stellar occultation by Vanth on 2014 March 01 returned one positive chord, 73 km in length (Braga-Ribas et al., 2017). Here, we place the tightest constraints to date on the size of Vanth, as well as any Vanth atmosphere and rings/debris in the Orcus-Vanth system. This was a particularly challenging occultation observation, with an unusually long deadtime at a critical moment at one our sites and an unexpected second star that complicated the event geometry. Neither Gaia data release showed the multiplicity of the star, and high-resolution images were required from a large telescope in order to correctly interpret the data. Our results highlight the need for high-cadence, low-deadtime instrumentation for occultation observations and the difficult nature of accurate TNO occultation predictions. Nonetheless, stellar occultations remain one of the best methods for obtaining information about distant bodies in the Solar System.

**Acknowledgments**

We thank Mike Brown and an anonymous reviewer for suggestions that improved the manuscript and ensured the correct interpretation of the data. We thank Steve Howell, Mark Everett, and the DSSI instrument team for acquiring and processing the speckle images. This work was partially supported by the South African National Research Foundation and by NASA grant #NNX15AJ82G to ASB. AAS was a visiting astronomer at the Infrared Telescope Facility, which is operated by the University of Hawai'i under contract NNH14CK55B with the National Aeronautics and Space Administration. This work makes use of observations from the LCOGT network which is owned and operated by Las Cumbres Observatory. The work includes observations obtained with the SARA Observatory, which is owned and operated by the Southeastern Association for Research in Astronomy (saraobservatory.org). These results made use of the Discovery Channel Telescope at Lowell Observatory. Lowell is a private, non-profit institution dedicated to astrophysical research and public appreciation of astronomy and operates the DCT in partnership with Boston University, the University of Maryland, the University of Toledo, Northern Arizona University and Yale University. The Large Monolithic Imager was built by Lowell Observatory using funds provided by the National Science Foundation (AST-1005313). This work has made use of data from the European Space Agency (ESA) mission

Gaia (https://www.cosmos.esa.int/gaia), processed by the Gaia Data Processing and Analysis Consortium (DPAC, https://www. cosmos.esa.int/web/gaia/dpac/consortium). Funding for the DPAC has been provided by national institutions, in particular the institutions participating in the Gaia Multilateral Agreement. This paper has made use of NASA's Astrophysics Data System.**Research data captions**

ResData1.IRTFMORIS.txt: Table containing midtimes of the exposures and normalized flux values for the IRTF MORIS light curve, near the time of the observed occultation.

ResData2.IRTFSpeX.txt: Table containing midtimes of the exposures and normalized flux values for the IRTF SpeX light curve, near the time of the observed occultation.

ResData3.ELP.txt: Table containing midtimes of the exposures and normalized flux values for the ELP light curve, near the time of the observed occultation.

**References**

Barucci, M.A., Merlin, F., Guilbert, A., de Bergh, C., Alvarez-Candal, A., Hainaut, O., Doressoundiram, A., Dumas, C., Owen, T., Coradini, A., 2008. Surface composition and temperature of the TNO Orcus. A&A 479, L13-L16.

Benedetti-Rossi, G., Sicardy, B., Buie, M.W., Ortiz, J.L., Vieira-Martins, R., Keller, J.M., Braga-Ribas, F., Camargo, J.I.B., Assafin, M., Morales, N., Duffard, R., Dias-Oliveira, A., Santos-Sanz, P., Desmars, J., Gomes-Júnior, A.R., Leiva, R., Bardecker, J., Bean, J.K., Jr., Olsen, A.M., Ruby, D.W., Sumner, R., Thirouin, A., Gómez-Muñoz, M.A., Gutierrez, L., Wasserman, L., Charbonneau, D., Irwin, J., Levine, S., Skiff, B., 2016. Results from the 2014 November 15th Multi-chord Stellar Occultation by the TNO (229762) 2007 UK126. AJ 152, 11.

Bérard, D., Sicardy, B., Camargo, J.I.B., Desmars, J., Braga-Ribas, F., Ortiz, J.-L., Duffard, R., Morales, N., Meza, E., Leiva Espinoza, R., Benedetti-Rossi, G., Assafin, M., Vieira-Martins, R., Colas, F., Dauvergne, J.-L., Kervella, P., Lecacheaux, J., Maquet, L., Vachier, F., Renner, S., Monrad, B., Sickafoose, A.A., Breytenbach, H., Genade, A., Beisker, W., Bath, K.-L., Bode, H.-J., Backes, M., Ivanov, V.D., Jehin, E., Gillon, M., Manfroid, J., Pollock, J., Tancredi, G., Roland, S., Salvo, R., Vanzi, L., Herald, D., Gault, D., Kerr, S., Pavlov, H., Hill, K.M., Bradshaw, J., Barry, M.A., Cool, A., Lade, B., Cole, A., Broughton, J., Newman, J., Horvat, R., Maybour, D., Giles, D., Davis, L., Paton, R.A., Loader, B., Pennell, A., Jaquiery, P.-D., Brilliant, S., Selman, F., Dumas, C., Herrera, C., Carraro, G., Monaco, L., Maury, A., Peyrot, A., Teng-Chuen-Yu, J.-P., Richichi, A., Irawati, P., De Witt, C., Schoenau, P., Prager, R., Colazo, C., Melia, R., Spagnotto, J., Blain, A., Alonso, S., Román, A., Santos-Sanz, P., Rizos, J.-L., Maestre, J.-L., Dunham, D., 2017. The structure of Chariklo's rings from stellar occultations. AJ 154, 21.

Bieler, A., Altwegg, K., Balsiger, H., Bar-Nun, A., Berthelier, J.-J., Boshsler, P., Briois, C., Calmonte, U., Combi, M., De Keyser, J., van Dishoeck, E.F., Fiethe, B., Fuselier, S.A., Gasc, S., Gombosi, T.I., Hansen, K.C., Hässig, M., Jäckel, A., Kopp, E., Korth, A., Le Roy, L., Mall, U., Maggiolo, R.,

Marty, B., Mousis, O., Owen, T., Rème, H., Rubin, M., Sémon, T., Tzou, C.-Y., Waite, J.H., Walsh, C., Wurz, P., 2015. Abundant molecular oxygen in the coma of comet 67P/Churyumov–Gerasimenko. Nature 526, 678-681.

Bosh, A.S., Person, M.J., Levine, S.E., Zuluaga, C.A., Zangari, A.M., Gulbis, A.A.S., Schaefer, G., Dunham, E.W., Babcock, B.A., Pasachoff, J.M., Rojo, P., Servajean, E., Forster, F., Oswalt, T., Batcheldor, D., Bell, D., Bird, P., Fey, D., Fulwider, T., Geisert, E., Hastings, D., Keuhler, C., Mizusawa, T., Solenski, P., Watson, B., 2015. The State of Pluto's Atmosphere in 2012-2013. Icarus 246, 237-246.

Braga-Ribas, F., Sicardy, B., Ortiz, J.L., Snodgrass, C., Roques, F., Vieira Martins, R., Camargo, J.I.B., Assafin, M., Duffard, R., Jehin, E., Pollock, J., Leiva, R., Emilio, M., Machado, D.I., Colazo, C., Lellouch, E., Skottfelt, J., Gillon, M., Ligier, N., Maquet, L., Benedetti-Rossi, G., Ramos Gomes Jr., A., Kervella, P., Monteiro, H., Sfair, R., El Moutamid, M., Tancredi, G., Spagnotto, J., Maury, A., Morales, N., Gil-Hutton, R., Roland, S., Ceretta, A., Gu, S.-h., Wang, X.-b., Harpsoe, K., Rabus, M., Manfroid, J., Opitom, C., Vanzi, L., Mehret, L., Lorenzini, L., Schneiter, E.M., Melia, R., Lecacheaux, J., Colas, F., Vachier, F., Widemann, T., Almenares, L., Sandness, R.G., Char, F., Perez, V., Lemos, P., Martinez, N., Jorgensen, U.G., Dominik, M., Roig, F., Reichart, D.E., LaCluyze, A.P., Haislip, J.B., Ivarsen, K.M., Moore, J.P., Frank, N.R., Lambas, D.G., 2014. A ring system detected around the Centaur (10199) Charliko. Nature 508, 72-75.

Braga-Ribas, F., Vachier, F., Camargo, J.I.B., Desmars, J., Sicardy, B., Vieira-Martins, R., Assafin, M., Benedetti-Rossi, G., Dias-Oliveira, A., Murakami, Y., Lecacheaux, J., 2017. Stellar Occultations by TNOs: Probing Rings, Surface, and Satellites, Asteroids, Comets, Meteors (ACM2017), Montevideo, Uruguay.

Brown, M.E., Butler, B.J., 2018. Medium-sized satellites of large Kuiper Belt objects. AJ submitted.

Brown, M.E., Ragozzine, D., Stansberry, J., Fraser, W.C., 2010. The Size, Density, and Formation of the Orcus-Vanth System in the Kuiper Belt. AJ 139, 5.

Bus, S.J., Buie, M.W., Schleicher, D.G., Hubbard, W.B., Marcialis, R.L., Hill, R., Wasserman, L.H., Spencer, J.R., Millis, R.L., Franz, O.G., Bosh, A.S., Dunham, E.W., Ford, C.H., Young, J.W., Elliot, J.L., Meserole, R., Olkin, C.B., McDonald, S.W., Foust, J.A., Sopata, L.M., Bandyopadhyay, R.M., 1996. Stellar occultation by 2060 Chiron. Icarus 123, 478-490.

Carry, B., Hestroffer, D., DeMeo, F.E., Thirouin, A., Berthier, J., Lacerda, P., Sicardy, B., Doressoundiram, A., Dumas, C., Farrelly, D., Muller, T.G., 2011. Integral-field spectroscopy of (90482) Orcus-Vanth. A&A 534.

Coppejans, R., Gulbis, A.A.S., Kotze, M.M., Coppejans, D.L., Worters, H.L., Woudt, P.A., Whittal, H.B., Cloete, J., Fourie, P., 2013. Characterizing and Commissioning the Sutherland High-speed Optical Cameras (SHOC). PASP 125, 976-988.

de Bergh, C., Delsanti, A., Tozzi, G.P., Dotto, E., Doressoundiram, A., Barucci, M.A., 2005. The surface of the transneptunian object 90482 Orcus. A&A 437, 1115-1120.

Delsanti, A., Merlin, F., Guilbert-Lepoutre, A., Bauer, J., Yang, B., Meech, K.J., 2010. Methane, ammonia, and their irradiation products at the surface of an intermediate-size KBO?. A portrait of Plutino (90482) Orcus. A&A 540, 15.

Duffard, R., Ortiz, J.L., Thirouin, A., Santos-Sanz, P., Morales, N., 2009. Transneptunian objects and Centaurs from light curves. A&A 505, 12.

Elliot, J., Zuluaga, C.A., Person, M.J., Adams, E.R., Lockhart, M., Zangari, A.M., Bosh, A.S., Gulbis, A.A.S., Levine, S.E., Sheppard, S.S., Dunham, E.W., Bright, L., Souza, S.P., Pasachoff, J.M.,

Babcock, B.A., Ryan, W.H., Ryan, E.V., 2009. The MIT Program for Predicting Stellar Occultations by Kuiper Belt Objects, AAS DPS Meeting, p. id. 62.09.

Elliot, J.L., Olkin, C.B., Dunham, E.W., Ford, C.H., Gilmore, D.K., Kurtz, D., Lazzaro, D., Rank, D.M., Temi, P., Bandyopadhyay, R.M., Barroso, J., Barucci, A., Bosh, A.S., Buie, M.W., Bus, S.J., Dahn, C.C., Foryta, D.W., Hubbard, W.B., Lopes, D.F., Marcialis, R.L., McDonald, S.W., Millis, R.L., Reitsema, H., Schleicher, D.G., Sicardy, B., Stone, R.P.S., Wasserman, L.H., 1995. Jet-like features near the nucleus of 2060 Chiron. Nature 373, 46-49.

Elliot, J.L., Person, M.J., Zuluaga, C.A., Bosh, A.S., Adams, E.R., Brothers, T.C., Gulbis, A.A.S., Levine, S.E., Lockhart, M., Zangari, A.M., Babcock, B.A., DuPre´, K., Pasachoff, J.M., Souza, S.P., Rosing, W., Secrest, N., Bright, L., Dunham, E.W., Sheppard, S.S., Kakkala, M., Tilleman, T., Berger, B., Briggs, J.W., Jacobson, G., Valleli, P., Volz, B., Rapoport, S., Hart, R., Brucker, M., Michel, R., Mattingly, A., Zambrano-Marin, L., Meyer, A.W., Wolf, J., Ryan, E.V., Ryan, W.H., Morzinski, K., Grigsby, B., Brimacombe, J., Ragozzine, D., Montano, H.G., Gilmore, A., 2010. Size and albedo of Kuiper belt object 55636 from a stellar occultation. Nature 465, 897-900.

Fornasier, S., Doressoundiram, A., Tozzi, G.P., Barucci, M.A., Boehnhardt, H., de Bergh, C., Delsanti, A., Davies, J., Dotto, E., 2004. ESO Large Program on physical studies of Trans-Neptunian objects and Centaurs: Final results of the visible spectrophotometric observations. A&A 421, 353-363.

Fornasier, S., Lellouch, E., Muller, T., Santos-Sanz, P., Panuzzo, P., Kiss, C., Lim, T., Mommert, M., Bockelee-Morvan, D., Vilenius, E., Stansberry, J., Tozzi, G.P., Mottola, S., Delsanti, A., Crovisier, J., Duffard, R., Henry, F., Lacerda, P., Barucci, A., Gicquel, A., 2013. TNOs are cool: A survey of the trans-Neptunian region, VII. Combined Herschel PACS and SPIRE observations of 9 bright targets at 70-500 um. A&A 555.

French, R.G., Gierasch, P.J., 1976. Diffraction calculation of occultation light curves in the presence of an isothermal atmosphere. AJ 81, 445-451.

Galiazzo, M., de la Fuente Marcos, C., de la Fuente Marcos, R., Carraro, G., Maris, M., Montalto, M., 2016. Photometry of Centaurs and trans-Neptunian objects: 2060 Chiron (1977 UB), 10199 Chariklo (1997 CU26), 38628 Huya (2000 EB173), 28978 Ixion (2001 KX76), and 90482 Orcus (2004 DW). Astrophysics and Space Science 361, 15.

Gulbis, A.A.S., Bus, S.J., Elliot, J.L., Rayner, J.T., Stahlberger, W.E., Rojas, F.E., Adams, E.R., Person, M.J., Chung, R., Tokunaga, A.T., Zuluaga, C.A., 2011. First Results from the MIT Optical Rapid Imaging System (MORIS) on the IRTF: A Stellar Occultation by Pluto and a Transit by Exoplanet XO-2b. PASP 123, 461-469.

Gulbis, A.A.S., Elliot, J.L., Person, M.J., Adams, E.R., Babcock, B.A., Emilio, M., Gangestad, J.W., Kern, S.D., Kramer, E.A., Osip, D.J., Pasachoff, J.M., Souza, S.P., Tuvikene, T., 2006. Charon's radius and atmospheric constraints from observations of a stellar occultation. Nature 439, 48-51.

Horch, E.P., Veillette, D.R., Baena Gallé, R., Shah, S.C., O'Rielly, G.V., van Altena, W.F., 2009. Observations of Binary Stars with the Differential Speckle Survey Instrument. I. Instrument Description and First Results. AJ 137, 10.

Howell, S.B., Everett, M.E., Sherry, W., Horch, E., Ciardi, D.R., 2011. Speckle Camera Observations for the NASA Kepler Mission Follow-up Program. ApJ 142, 9.

Keel, W.C., Oswalt, T., Mack, P., Henson, G., Hillwig, T., Batcheldor, D., Berrington, R., De Pree, C., Hartmann, D., Leake, M., Licandro, J., Murphy, B., Webb, J., Wood, M.A., 2017. The Remote Observatories of the Southeastern Association for Research in Astronomy (SARA). PASP 129, 12.

Lim, T.L., Stansberry, J., Müller, T.G., Mueller, M., Lellouch, E., Kiss, C., Santos-Sanz, P., Vilenius, E., Protopapa, S., Moreno, R., Delsanti, A., Duffard, R., Fornasier, S., Groussin, O., Harris, A.W., Henry, F., Horner, J., Lacerda, P., Mommert, M., Ortiz, J.L., Rengel, M., Thirouin, A., Trilling, D., Barucci, A., Crovisier, J., Doressoundiram, A., Dotto, E., Gutiérrez Buenestado, P.J., Hainaut, O., Hartogh, P., Hestroffer, D., Kidger, M., Lara, L., Swinyard, B.M., Thomas, N., 2010. "TNOs are Cool": A survey of the trans-Neptunian region . III. Thermophysical properties of 90482 Orcus and 136472 Makemake. A&A 518, 5.

Meech, K.J., Svoren, J., 2004. Using Cometary Activity to Trace the Physical and Chemical Evolution of Cometary Nuclei, in: M.C. Festou, H.U.K., H. A. Weaver (Ed.), Comets II. University of Arizona Press, Tucson, pp. 317-335.

Millis, R.L., Wasserman, L.H., Franz, O.G., Nye, R.A., Elliot, J.L., Dunham, E.W., Bosh, A.S., Young, L.A., Slivan, S.M., Gilmore, A.C., Kilmartin, P.M., Allan, W.H., Watson, R.D., Dieters, S.W., Hill, K.M., Giles, A.B., Blow, G., Priestly, J., Kissling, W.M., Walker, W.S.G., Marino, B.F., Dix, D.G., Page, A.A., Ross, J.E., Avey, H.P., Hickey, D., Kennedy, H.D., Mottram, K.A., Moyland, G., Murphy, T., Dahn, C.C., Klemola, A.R., 1993. Pluto's radius and atmosphere: Results from the entire 9 June 1988 occultation data set. Icarus 105, 282-297.

Mousis, O., Lunine, J.I., Mandt, K.E., Schindhem, E., Weaver, H.A., Stern, S.A., Waite, J.H., Gladstone, R., Moudens, A., 2013. On the possible noble gas deficiency of Pluto's atmosphere. Icarus 225, 856-861.

Ortiz, J.L., Duffard, R., Pinilla-Alonso, N., Alvaraz-Candal, A., Santos-Sanz, P., Morales, N., Fernandez-Valenzuela, E., Licandro, J., Campo Bagatin, A., Thirouin, A., 2015. Possible ring material around centaur (2060) Chiron. A&A 576, A18.

Ortiz, J.L., Santos-Sanz, P., Sicardy, B., Benedetti-Rossi, G., Bérard, D., Morales, N., Duffard, R., Braga-Ribas, F., Hopp, U.R., C., Nascimbeni, V., Marzari, F., Granata, V., Pál, A., Kiss, C., Pribulla, T., Komžík, R., Hornoch, K., Pravec, P., Bacci, P., Maestripieri, M., Nerli, L., Mazzei, L., Bachini, M., Martinelli, F., Succi, G., Ciabattari, F., Mikuz, H., Carbognani, A., Gaehrken, B., Mottola, S., Hellmich, S., Rommel, F.L., Fernández-Valenzuela, E., Bagatin, A.C., Cikota, S., Cikota, A., Lecacheux, J., Vieira-Martins, R., Camargo, J.I.B., Assafin, M., Colas, F., Behrend, R., Desmars, J., Meza, E., Alvarez-Candal, A., Beisker, W., Gomes-Junior, A.R., Morgado, B.E., Roques, F., Vachier, F., Berthier, J., Mueller, T.G., Madiedo, J.M., Unsalan, O., Sonbas, E., Karaman, N., Erece, O., Koseoglu, D.T., Ozisik, T., Kalkan, S., Guney, Y., Niaei, M.S., Satir, O., Yesilyaprak, C., Puskullu, C., Kabas, A., Demircan, O., Alikakos, J., Charmandaris, V., Leto, G., Ohlert, J., Christille, J.M., Szakáts, R., Farkas, A.T., Varga-Verebélyi, E., Marton, G., Marciniak, A., Bartczak, P., Santana-Ros, T., Butkiewicz-Bąk, M., Dudziński, G., Alí-Lagoa, V., Gazeas, K., Tzouganatos, L., Paschalis, N., Tsamis, V., Sánchez-Lavega, A., Pérez-Hoyos, S., Hueso, R., Guirado, J.C., Peris, V., Iglesias-Marzoa, R., 2017. The size, shape, density and ring of the dwarf planet Haumea from a stellar occultation. Nature 550, 219-223.

Person, M.J., Elliot, J.L., Gulbis, A.A.S., Zuluaga, C.A., Babcock, B.A., McKay, A.J., Pasachoff, J.M., Souza, S.P., Hubbard, W.B., Kulesa, C.A., McCarthy, D.W., Kern, S.D., Levine, S.E., Bosh, A.S., Ryan, E.V., Ryan, W.H., Meyer, A., Wolf, J., 2008. Waves in Pluto's Upper Atmosphere. AJ 136, 1510-1518.

Pfuller, E., 2016. Fast EMCCD Cameras for the Optical Characterization of the SOFIA Observatory and its Telescope Subsystems. Universitat Stuttgart.

Rayner, J.T., Toomey, D.W., Onaka, P.M., Denault, A.J., Stahlberger, W.E., Vacca, W.D., Cushing, M.C., Wang, S., 2003. SpeX: A medium-resolution 0.8-5.4 micron spectrograph and imager for the NASA Infrared Telescope Facility. PASP 115, 362-382.

Ruprecht, J.D., Bosh, A.S., Person, M.J., Bianco, F.B., Fulton, B.J., Gulbis, A.A.S., Bus, S.J., Zangari, A.M., 2015. 29 November 2011 Stellar Occultation by 2060 Chiron: Symmetric jet-like features. Icarus 252, 271-276.

Schaller, E.L., Brown, M., 2007. Volatile Loss and Retention on Kuiper Belt Objects. ApL 659, L61.

Schindler, K., Wolf, J., Bardecker, J., Olsen, A., Müeller, T., Kiss, C., Ortiz, J.L., Braga-Ribas, F., Camargo, J.I.B., Herald, D., Krabbe, A., 2017. Results from a triple chord stellar occultation and far-infrared photometry of the trans-Neptunian object (229762) 2007 UK126. A&A 600, 16.

Sheppard, S.S., 2007. Light Curves of Dwarf Plutonian Planets and Other Large Kupier Belt Objects: Their roatations, phase functions, and absolute magnitudes. AJ 134, 787-798.

Sicardy, B., Ortiz, J.L., Assafin, M., Jehin, E., Maury, A., Lellouch, E., Hutton, R.G., Braga-Ribas, F., Colas, F., Hestroffer, D., Lecacheux, J., Roques, F., Santos-Sanz, P., Widemann, T., Morales, N., Duffard, R., Thirouin, A., Castro-Tirado, A.J., Jelinek, M., Kubanek, P., Sota, A., Sanchez-Ramierez, R., Andrei, A.H., Camargo, J.I.B., da Silva Neto, D.N., Gomes, A.R., Martins, R.V., Gillon, M., Manfroid, J., Tozzi, G.P., Harlingten, C., Saravia, S., Behrend, R., Mottola, S., Melendo, E.G., Peris, V., Fabregat, J., Madiedo, J.M., Cuesta, L., Eibe, M.T., Ullán, A., Organero, F., Pastor, S., de Los Reyes, J.A., Pedraz, S., Castro, A., de La Cueva, I., Muler, G., Steele, I.A., Cebrián, M., Montañés-Rodríguez, P., Oscoz, A., Weaver, D., Jacques, C., Corradi, W.J.B., Santos, F.P., Reis, W., Milone, A., Emilio, M., Gutiérrez, L., Vázquez, R., Hernández-Toledo, H., 2011. A Pluto-like radius and a high albedo for the dwarf planet Eris from an occultation. Nature 478, 493-496.

Simon, S., Saur, J., Neubauer, F.M., Wennmacher, A., Dougherty, M.K., 2011. Magnetic signatures of a tenuous atmosphere at Dione. GRL 38, CiteID L15102.

Souza, S.P., Babcock, B.A., Pasachoff, J.M., Gulbis, A.A.S., Elliot, J.L., Person, M.J., Gangestad, J.W., 2006. POETS: Portable Occultation, Eclipse, and Transit System. PASP 118, 1550-1557.

Stansberry, J.A., Grundy, W., Brown, M., Cruikshank, D., Spencer, J., Trilling, D.E., Margot, J.-L., 2008. Physical Properties of Kuiper Belt and Centaur Objects: Constraints from Spitzer Space Telescope, in: Barucci, M.A., Boehnhardt, H., Cruikshank, D.P., Morbidelli, A. (Eds.), The Solar System beyond Neptune. The University of Arizona Press, Tucson, pp. 161-179.

Stetson, P.B., 1987. DAOPHOT: A computer program for crowded-field stellar photometry. PASP 99, 191 - 222.

Thirouin, A., Ortiz, J.L., Duffard, R., Santos-Sanz, P., Aceituno, F.J., Morales, N., 2010. Short-term variability of a sample of 29 trans-Neptunian objects and Centaurs. A&A 522, 43.

Throop, H.B., French, R.G., Shoemaker, K., Olkin, C.B., Ruhland, T.R., Young, L.A., 2015. Limits on Pluto's ring system from the June 12 2006 stellar occultation and implications for the New Horizons impact hazard. Icarus 246, 345-351.

Trujillo, C.A., Brown, M.E., Rabinowitz, D.L., Geballe, T.R., 2005. Near-infrared surface properties of the two intrinsically brightest minor planets: (90377) Sedna and (90482) Orcus. ApJ 627, 1057 - 1065.

van Belle, G.T., 1999. Predicting stellar angular sizes. PASP 111, 1515-1523.

Womack, M., Sarid, G., Wierzchos, K., 2017. CO and Other Volatiles in Distantly Active Comets. PASP 129, 20.

Young, E.F., French, R.G., Young, L.A., Ruhland, C.R., Buie, M.W., Olkin, C.B., Register, J., Shoemaker, K., Blow, G., Broughton, J., Christie, G., Gault, D., Lade, B., Natusch, T., 2008.

Vertical structure in Pluto's atmosphere from the 2006 June 12 stellar occultation. AJ 136, 1757-1769.

Young, L.A., Kammer, J.A., Steffl, A.J., Gladstone, G.R., Summers, M.E., Strobel, D.F., Hinson, D.P., Stern, S.A., Weaver, H.A., Olkin, C.B., Ennico, K., McComas, D.J., Cheng, A.F., Gao, P., Lavvas, P., Linscott, I.R., Wong, M.L., Yung, Y.L., Cunningham, N., Davis, M., Parker, J.W., Schindhelm, E., Siegmund, O.H.W., Stone, J., Retherford, K., Versteeg, M.H., 2017. Structure and Composition of Pluto's atmosphere from the New Horizons Solar Ultraviolet Occultation. Icarus 300, 174-199.